\documentclass[aps,pre,twocolumn,amsmath,amssymb]{revtex4}

\usepackage{graphicx}
\usepackage{enumitem}
\usepackage[usenames, dvipsnames]{color}
\usepackage{multirow,dcolumn}
\usepackage{newtxtext,newtxmath}

\begin{document}

\title{Objective measures for sentinel surveillance in network epidemiology}

\author{Petter Holme}
\email{holme@cns.pi.titech.ac.jp}
\affiliation{Institute of Innovative Research, Tokyo Institute of Technology, Nagatsuta-cho 4259, Midori-ku, Yokohama, Kanagawa, 226-8503, Japan}

\begin{abstract}
Assume one has the capability of determining whether a node in a network is infectious or not by probing them. Then problem of optimizing sentinel surveillance in networks is to identify the nodes to probe such that an emerging disease outbreak can be discovered early or reliably. Whether the emphasis should be on early or reliable detection depends on the scenario in question. We investigate three objective measures from the literature quantifying the performance of nodes in sentinel surveillance---the time to detection or extinction, the time to detection, and the frequency of detection. As a basis for the comparison, we use the susceptible-infectious-recovered model on static and temporal networks of human contacts. We show that, for some regions of parameter space, the three objective measures can rank the nodes very differently. This means sentinel surveillance is a class of problems, and solutions need to chose an objective measure for the particular scenario in question. As opposed to other problems in network epidemiology, we draw similar conclusions from the static and temporal networks. Furthermore, we do not find one type of network structure that predicts the objective measures---that depends both on the data set and the SIR parameter values.
\end{abstract}

\maketitle

\section{Introduction}

Infectious diseases are a big burden to public health. Their epidemiology is a topic where the step between the medical and theoretical sciences is not so far. Several concepts of mathematical epidemiology---like the basic reproductive number or core groups~\cite{giesecke,hethcote,andersonmay}---have entered the vocabulary of medical scientists. Traditionally, authors have modeled disease outbreaks in society by assuming any person have the same chance of meeting anyone else at any time. This is of course not realistic, and improving this point is the motivation for network epidemiology---epidemic simulations between people connected by a network~\cite{ps_rmp}. One can continue increasing the realism in the contact patterns by observing that the timing of contacts can also have structure capable of affecting the disease. Studying epidemics on time-varying contact structures is the basis of the emerging field of temporal network epidemiology~\cite{masuda_holme_rev,masuda_holme_intro,masuda_lambiotte,holme_modern}.

One of the most important questions in infectious disease epidemiology is to identify people, or in more general terms, units, that would get infected early and likely in an infectious outbreak. This is the \textit{sentinel surveillance} problem~\cite{Bajardirsif20120289,christakis_fowler_sentinels}. It is the aspect of node importance, that is the one most actively used in public health practice. Typically, it works by selecting some hospitals (clinics, cattle farms, etc.) to screen, or more frequently test, for a specific infection~\cite{publichealth}.

\begin{figure*}
\includegraphics[width=0.85\textwidth]{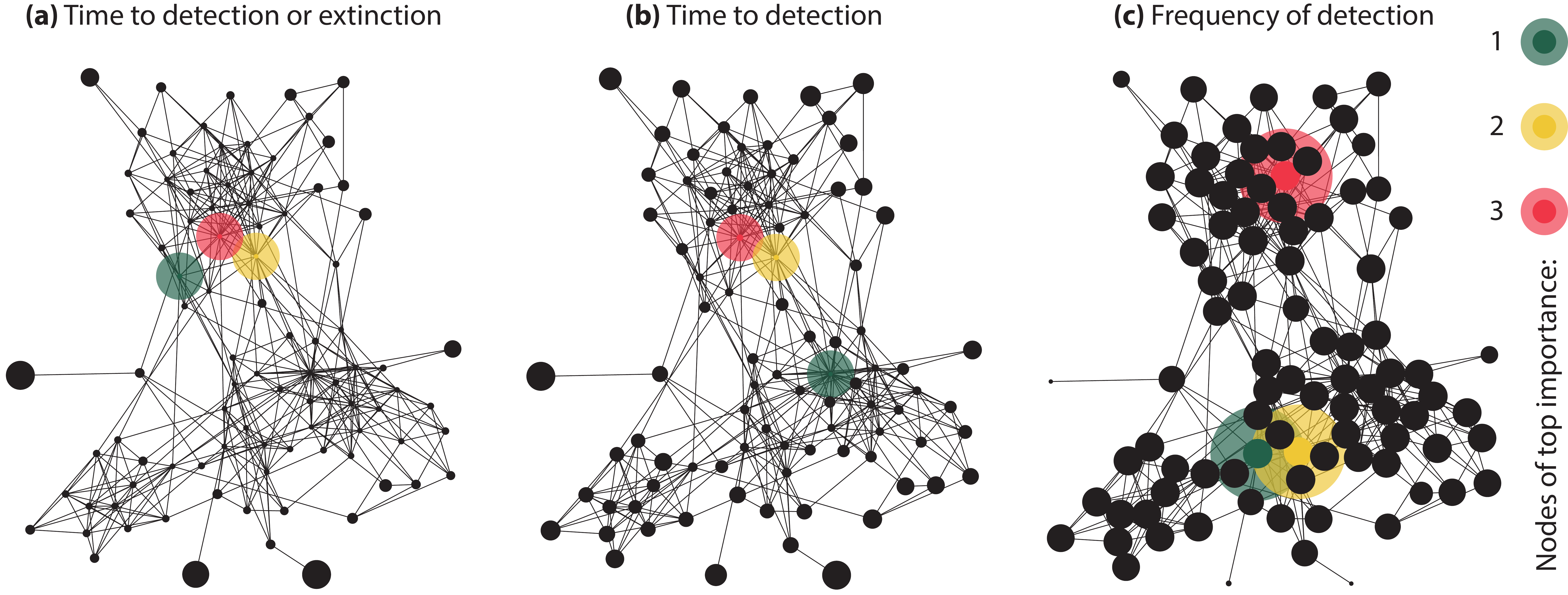}
\caption{(Color online) Objective measures for the \textit{Office} data with the infection rate $\beta = 1$: (a) the time to detection or extinction, (b) time to detection, (c) frequency of detection. The area of the circles are proportional to the respective objective measure. This means that in panel (c), the most important node is the largest, while in (a) and (b) it is the smallest. The three most important nodes of each panel are highlighted.}
\label{fig:office1}
\end{figure*}

Defining an objective measure---a quantity to be maximized or minimized---for sentinel surveillance is not trivial. It depends on the particular scenario one considers and the means of interventions at hand. If the goal for society is to detect as many outbreaks as possible, it makes sense to choose sentinels to maximize the fraction of detected outbreaks~\cite{Bajardirsif20120289}. If the objective rather is to discover outbreaks early, then one could choose sentinels that, if infected, are infected early~\cite{christakis_fowler_sentinels,bai}. Finally, if the objective is to stop the disease as early as possible, it makes sense to measure the time to extinction or detection (infection of a sentinel)~\cite{holme3f}. See Fig.~\ref{fig:office1} for an illustration. To restrict ourselves, we will focus on the case of one sentinel. If one has more than one sentinel, the optimal set will most likely not be the top nodes of a ranking according to the three measures above. Their relative position in the network also matter (they should not be too close to each other)~\cite{holme3f}.

In this paper, we study and characterize our three objective measures. We base our analysis on 38 empirical data sets of contacts between people. We analyze them both in temporal and static networks. The reason we use empirical contact data, rather than generative models, as the basis of this study is twofold. First, there are so many possible structures and correlations in temporal networks that one cannot tune them all in models~\cite{holme_modern}. It is also hard to identify the most important structures for a specific spreading phenomenon~\cite{holme_modern}. Second, studying empirical networks makes this paper---in addition to elucidating the objective measures of sentinel surveillance---a study of human interaction. We can classify data sets with respect how the epidemic dynamics propagate on them. As mentioned above, in practical sentinel surveillance, the network in question is rather one of hospitals, clinics or farms. One can however also think of sentinel surveillance of individuals, where high-risk individuals would be tested extra often for some diseases.

In the remainder of the paper, we will describe the objective measures, the structural measures we use for the analysis, the data sets, and present the analysis itself. We will primarily focus on the relation between the measures, secondarily on the structural explanations for our observations.

\section{Methods}

\subsection{Objective measures}

\subsubsection{Time to detection or extinction}

Assume that the objective of society is to end outbreaks as soon as possible. If an outbreak dies by itself, that is fine. Otherwise, one would like to detect it so it could be mitigated by interventions. In this scenario, a sensible objective measure would be the time for a disease to either go extinct or be detected by a sentinel---\textit{the time to detection or extinction} $t_x$~\cite{holme3f}.

\subsubsection{Time to detection}

Suppose that, in contrast to the situation above, the priority is not to save society from the epidemics as soon as possible, but just to detect outbreaks fast. This could be the case if one would want to get a chance to isolate a pathogen, or start producing a vaccine, as early as possible, maybe to prevent future outbreaks of the same pathogen at the earliest possibility. Then one would seek to minimize the time for the outbreak to be detected conditioned on the fact that it is detected---\textit{the time to detection} $t_d$.

\subsubsection{Frequency of detection}

For the time to detection, it does not matter how likely it is for an outbreak to reach a sentinel. If the objective is to detect as many outbreaks as possible, the corresponding measure should be the expected frequency of outbreaks to reach a node---the \textit{frequency of detection} $f_d$.

Note that for this measure, a large value means the node is a good sentinel, whereas for $t_x$ and $t_d$, a good sentinel has a low value. This means that when we correlate the measures, a similar ranking between $t_x$ and $f_d$ or $t_d$ and $f_d$ yields a negative correlation coefficient. Instead of considering the inverse times, or similar, we keep this feature and urge the reader to keep this in mind.

\begin{figure*}
\includegraphics[width=0.8\textwidth]{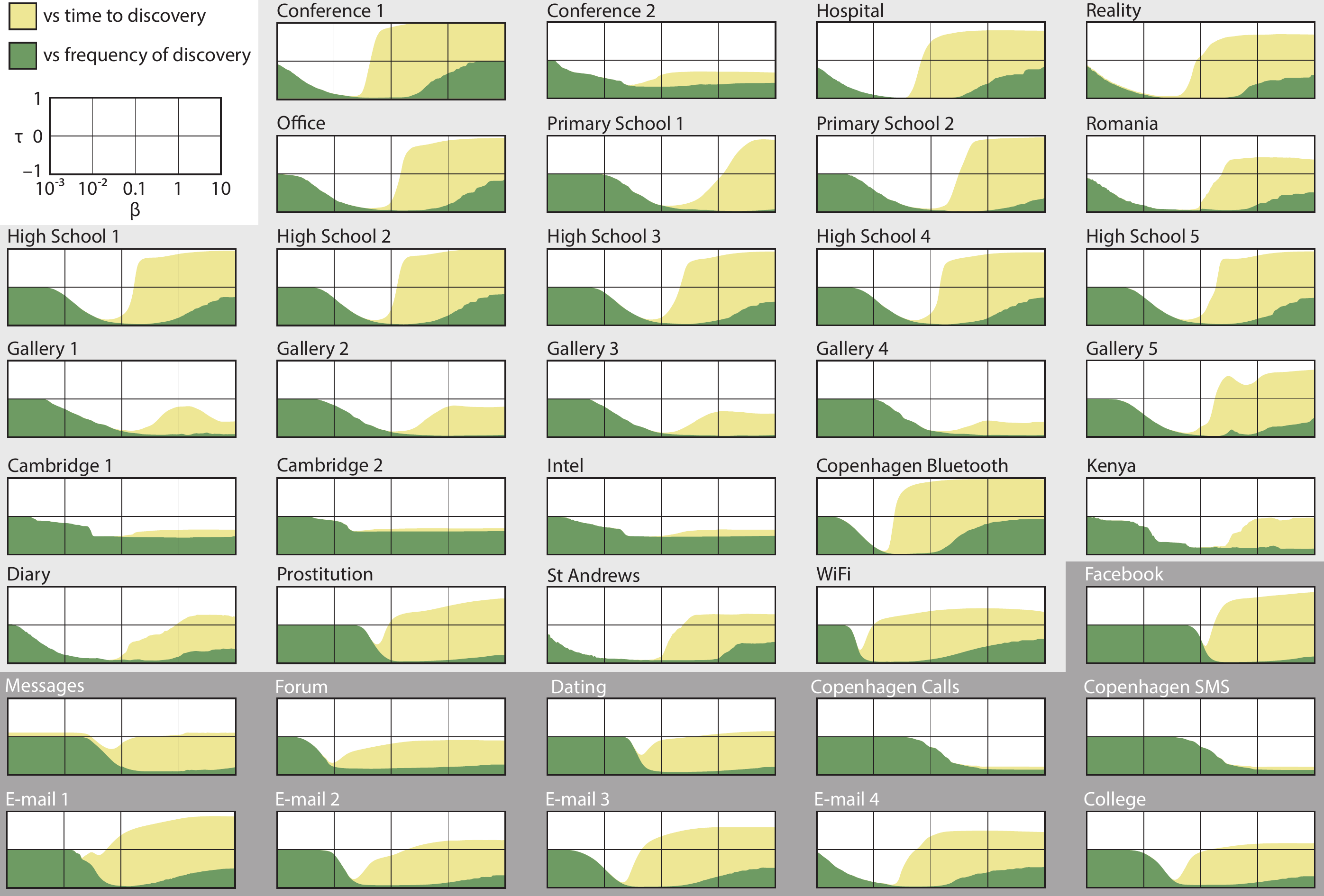}
\caption{(Color online) Kendall-$\tau$ correlation between the different objective measures for the static networks. For every data set, we show the correlations between time to detection or extinction the other two objective measures. Since $\tau$ for correlations with the frequency of detection is never larger than the correlations with the time to detection, we can highlight the curves by coloring the area underneath (without any point being covered). The $\beta$-axis is logarithmic.}
\label{fig:mcorr_stat}
\end{figure*}

\begin{figure*}
\includegraphics[width=0.8\textwidth]{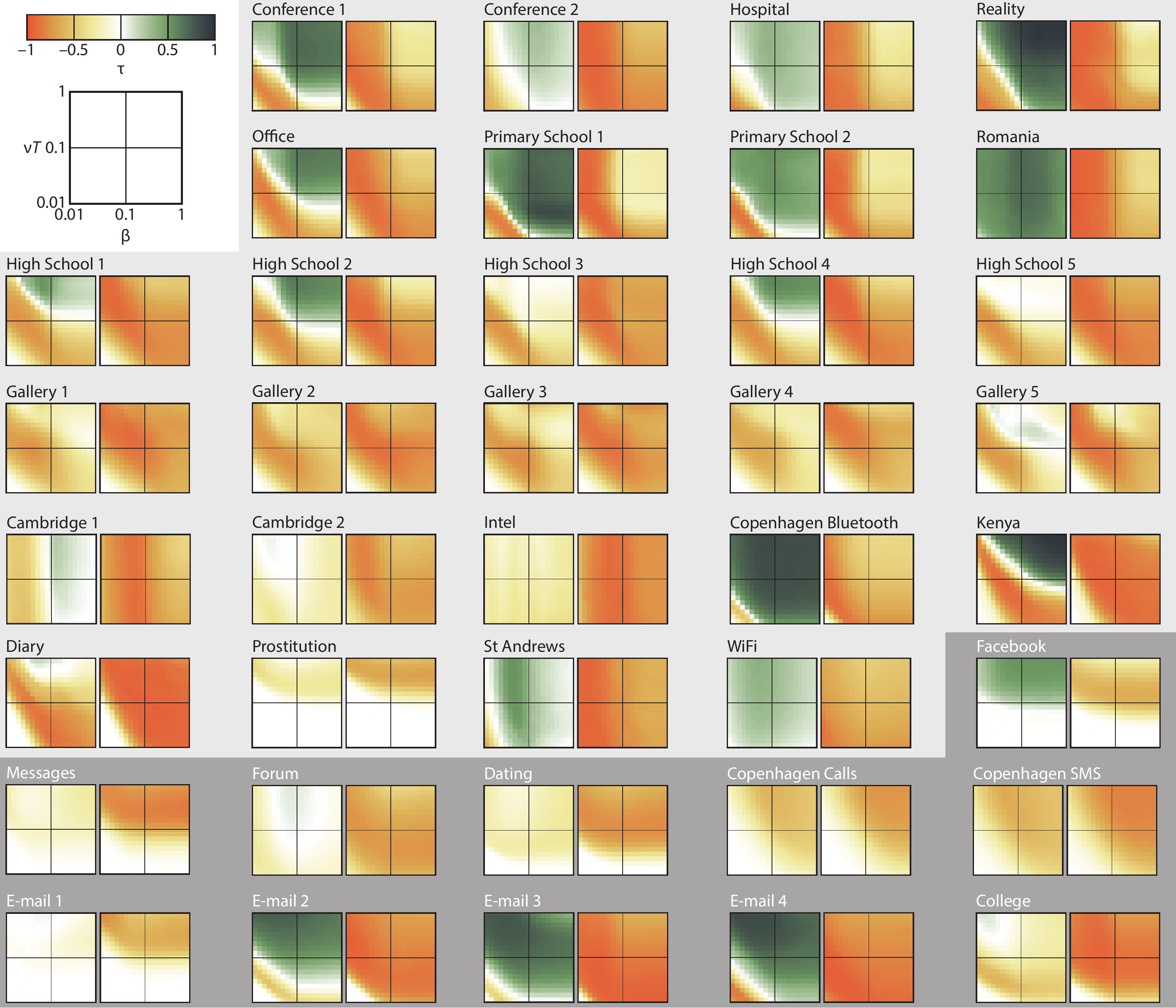}
\caption{(Color online) Correlation between the different objective measures for the temporal networks. For every data set, the left panel shows the Kendall-$\tau$ for correlations between time to detection or extinction and time to detection, whereas the right panel shows the corresponding values for time to detection or extinction and frequency of detection. The axes are logarithmic.}
\label{fig:mcorr_temp}
\end{figure*}

\begin{figure}
\includegraphics[width=0.9 \columnwidth]{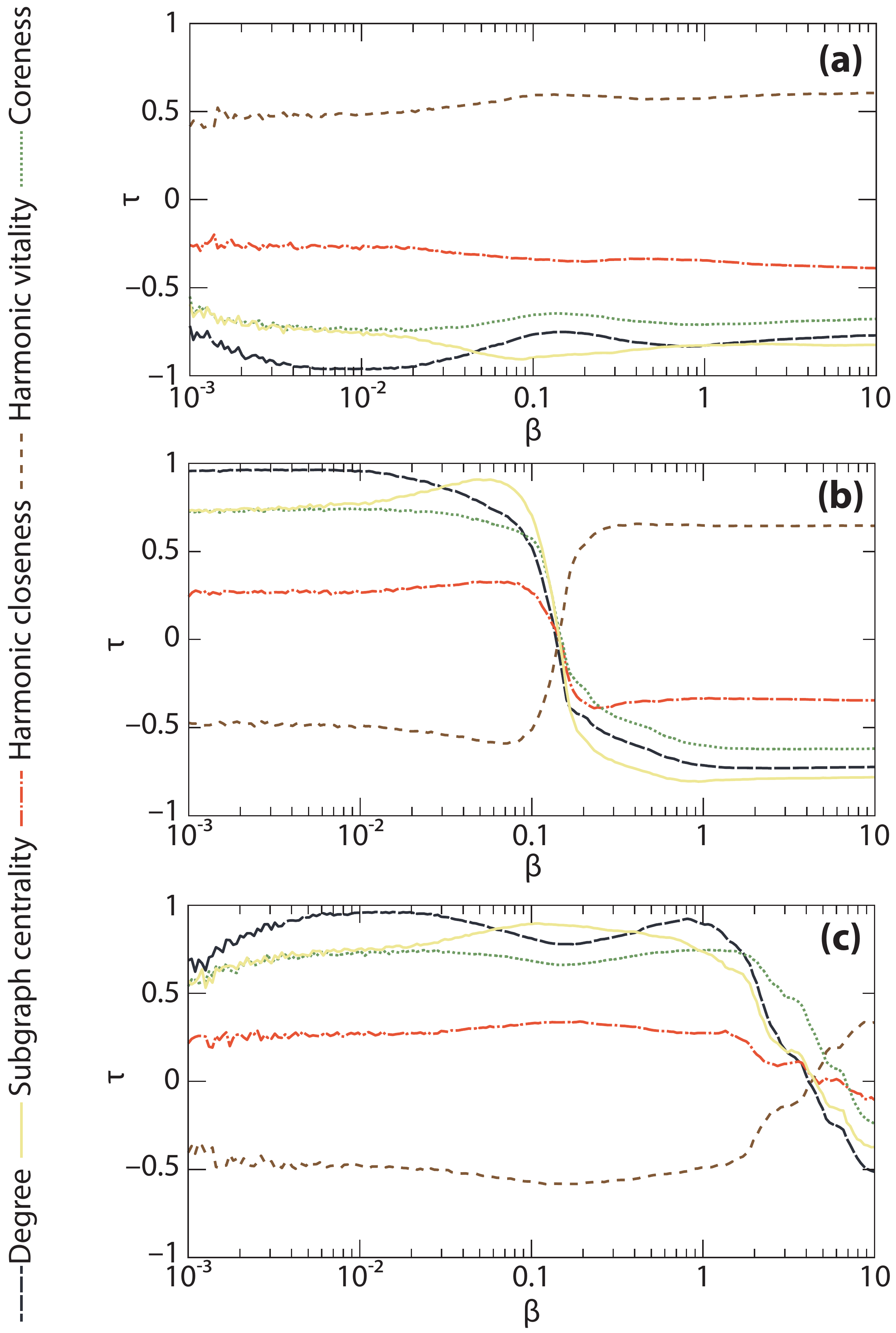}
\caption{(Color online) Correlations between the three objective measures and various quantities describing the static network structure for the \textit{Office} data set. Panel (a) shows results for the time to detection or extinction, (b) shows results for the time to detection, (c) shows results for the frequency of detection.}
\label{fig:office_stat_corr}
\end{figure}

\begin{figure*}
\includegraphics[width=\textwidth]{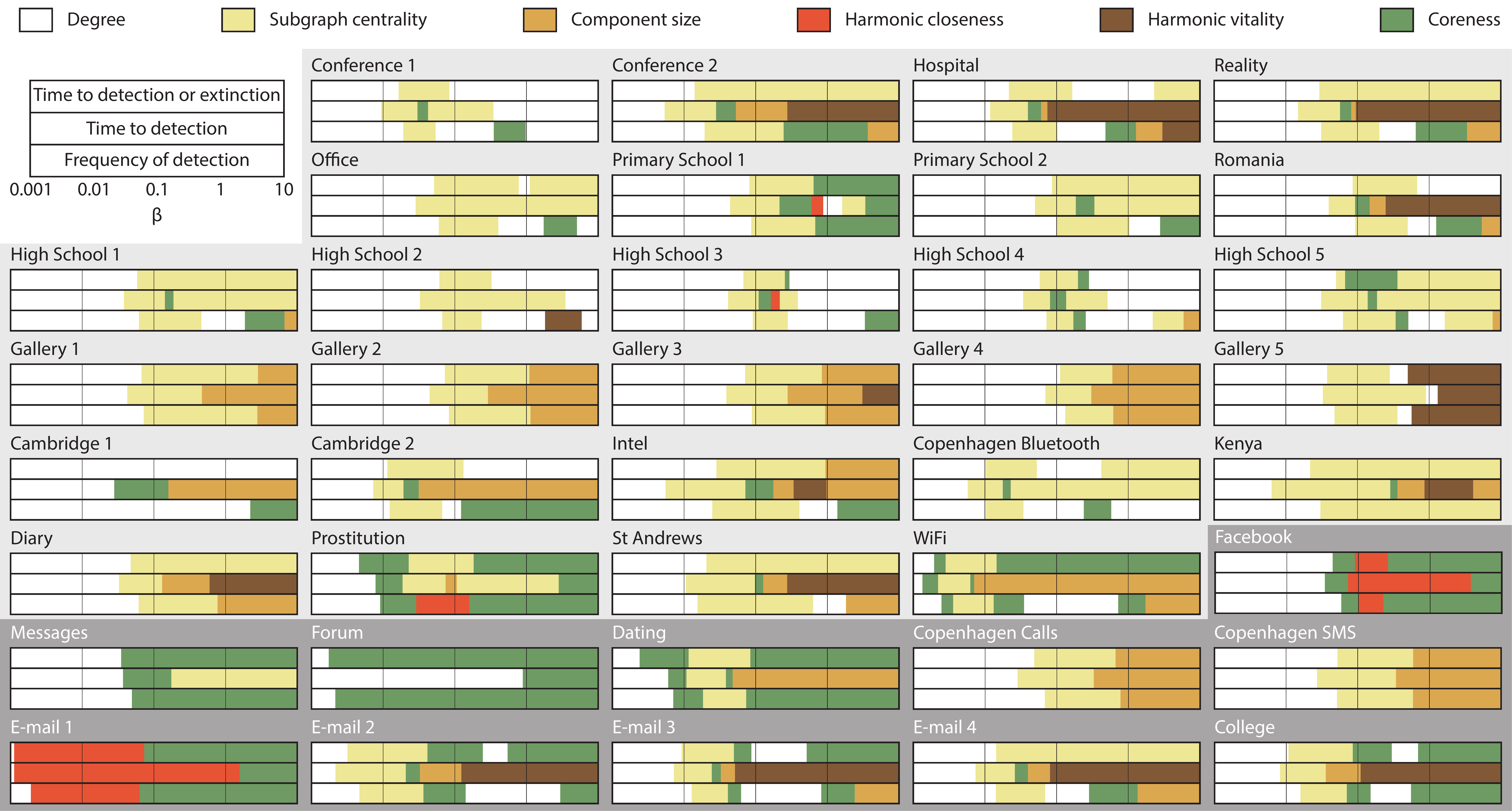}
\caption{(Color online) The strongest correlation between the three objective measures and various measures of the position of nodes for the static networks. Lighter shaded background are data sets of human proximity, the darker background figures indicates data of human communication. The $\beta$-axis is logarithmic.}
\label{fig:stat_struct}
\end{figure*}

\begin{figure}
\includegraphics[width=0.9\columnwidth]{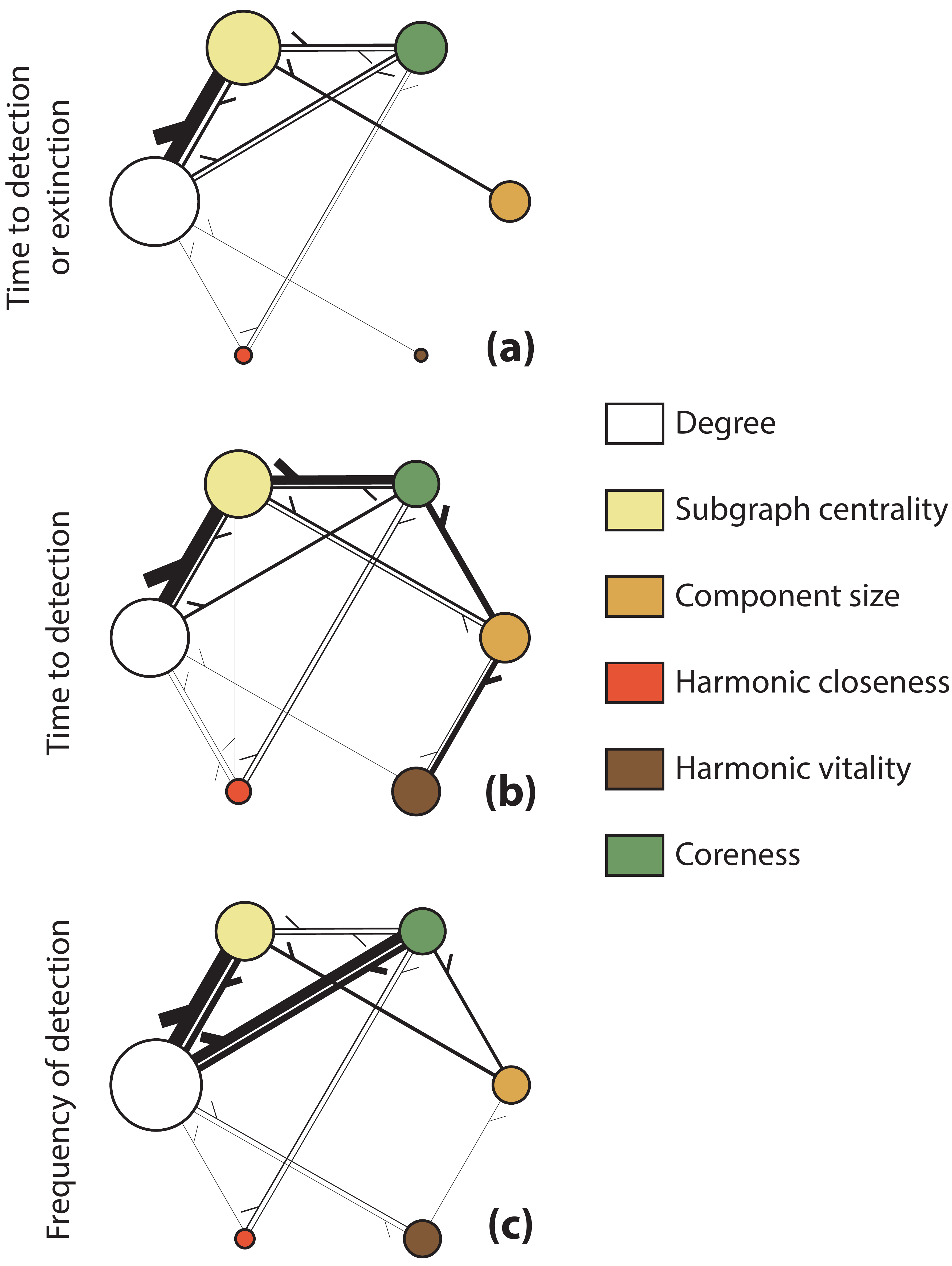}
\caption{(Color online) Transition graphs. The area of the circles correspond to the frequency of the structural measure in Fig.~\ref{fig:stat_struct}. The width of the lines are proportional to how many times one measure is succeeded by another as $\beta$ increases.}
\label{fig:transitions}
\end{figure}

\subsection{Reducing temporal networks to static networks}

There are many possible ways to reduce our empirical temporal networks to static networks. The simplest method would be to just include a link between any pair of nodes that has at least one contact during the course of the data set. This would however make some of the networks so dense that the static network structure of the node-pairs most actively in contact would be obscured. For our purpose, we primarily want our network to span many types of network structures that can impact epidemics.  There are some elaborate methods proposed for this purpose that, however, assumes extra knowledge about the epidemics~\cite{holme:compbio}. Without such extra knowledge, the best option is to threshold the weighted graph where an edge $(i,j)$ means that $i$ and $j$ had more than $\theta$ contacts in the data set. In this work, we assume that we do not know what the per-contact transmission probability $\beta$ is (this would anyway depend on both the disease and precise details of the interaction). Rather we scan through a very large range of $\beta$ values. Since we anyway to that, there is no need either to base the choice of $\theta$ on some epidemiological argument, or to rescale $\beta$ after the thresholding. Note that the rescaled $\beta$ would be a non-linear function of the number of contacts between $i$ and $j$. (Assuming no recovery, for an isolated link with $\nu$ contacts, the transmission probability is $1-(1-\beta)^\nu$.) For our purpose the only thing we need is that the rescaled $\beta$ is a monotonous function of $\beta$ for the temporal network 
(which is true). To follow a simple principle, we omit all links with a weight less than the median weight $\theta$.

\subsection{Disease simulations}

We simulate disease spreading by the SIR dynamics---the canonical model for diseases that gives immunity upon recovery~\cite{andersson,hethcote,andersson}. For static networks, we use the standard Markovian version of the SIR model~\cite{bartlett}. I.e.\  we assume that diseases spread over links between susceptible and infectious nodes the infinitesimal time interval ${\rm d}t$ with a probability $\beta \, {\rm d} t$. Then, an infectious node recovers after a time that is exponentially distributed with average $1/\nu$. The parameters $\beta$ and $\nu$ are called \textit{infection rate} and \textit{recovery rate}, respectively. We can, without loss of generality, put $\nu=1/T$ (where $T$ is the duration of the sampling). For other $\nu$ values, the ranking of the nodes would be the same (but the values of the $t_x$ and $t_d$ would be rescaled by a factor $\nu$). We will scan an exponentially increasing progression of 200 values of $\beta$, from $10^{-3}$ to $10$. The code for the disease simulations can be downloaded~\cite{github}.

For the temporal networks, we use a definition as close as possible to the one above. We assume an exponentially distributed duration of the infectious state with mean $1/\nu$. We assume a contact between an infectious and susceptible node results in a new infection with probability $\beta$. In the case of temporal networks, one cannot reduce the problem to one parameter. Like for static networks, we sample the parameter values in exponential sequences in the intervals $0.01\leq\beta\leq 1$ and $0.01\leq\nu/T\leq 1$ respectively. For temporal networks, with our interpretation of a contact, $\beta>1$ makes no sense, which explains the upper limit. Furthermore, since temporal networks usually are effectively sparser (in terms of the number of possible infection events per time), the smallest $\beta$ values will give similar results, which is the reason for the higher cutoff in this case.

For both temporal and static networks, we assume the outbreak starts at one randomly chosen node. Analogously, in the temporal case we assume the disease is introduced with equal probability at any time throughout the sampling period. For every data set and set of parameter values, we sample $10^7$ runs of epidemic simulations.

\begin{table*}
\caption{\label{tab:data}Basic statistics of the empirical temporal networks. $N$ is the number of nodes; $C$ is the number of contacts; $T$ is the total sampling time; $\Delta t$ is the time resolution of the data set, $M$ is the number of links in the projected and thresholded static networks, and $\theta$ is the threshold.}
\begin{ruledtabular}
\begin{tabular}{l|rrrrrrl}
Data set & $N$ & $C$ & $T$ & $\Delta t$ & $M$ & $\theta$ &Ref.\\ \hline
\textit{Conference 1} & 113 & 20,818 & 2.50d & 20s & 1,321 & 2&\cite{conference}\\
\textit{Conference 2} & 198 & 327,333 & 2.95d & 20s & 775 & 75 &\cite{cambridge} \\
\textit{Hospital} & 75 & 32,424 & 4.02d & 20s & 582 & 8&\cite{hospital} \\
\textit{Reality} & 63 & 26,260 & 8.63h & 5s & 421 & 3&\cite{reality} \\
\textit{Office} & 92 & 9,827 & 11.4d & 20s & 389 & 3&\cite{office} \\
\textit{Primary School 1} & 236 & 60,623 & 8.64h & 20s & 299 & 3&\cite{school} \\
\textit{Primary School 2} & 238 & 65,150 & 8.58h & 20s & 257 & 3&\cite{school} \\
\textit{Romania} & 42 & 1,748,401 & 62.8d & 1m & 128 & 61 &\cite{roman} \\
\textit{High School 1} & 312 & 28,780 & 4.99h & 20s & 1,385 &2 &\cite{hschool} \\
\textit{High School 2} & 310 & 47,338 & 8.99h & 20s & 1,601 & 2&\cite{hschool} \\
\textit{High School 3} & 303 & 40,174 & 8.99h & 20s & 1,096 &3&\cite{hschool} \\
\textit{High School 4} & 295 & 37,279 & 8.99h & 20s & 1,363 & 2&\cite{hschool} \\
\textit{High School 5} & 299 & 34,937 & 8.99h & 20s & 1,298 & 2&\cite{hschool} \\
\textit{Gallery 1} & 200 & 5,943 & 7.80h & 20s & 398 & 2&\cite{gallery} \\
\textit{Gallery 2} & 204 & 6,709 & 8.05h & 20s & 393 & 2&\cite{gallery} \\
\textit{Gallery 3} & 186 & 5,691 & 7.39h & 20s & 362 & 2&\cite{gallery} \\
\textit{Gallery 4} & 211 & 7,409 & 8.01h & 20s & 294 & 2&\cite{gallery} \\
\textit{Gallery 5} & 215 & 7,634 & 5.61h & 20s & 967 & 1&\cite{gallery} \\
\textit{Cambridge 1} & 186 & 3,853,714 & 6.07d & 20s & 180 & 1,312 &\cite{cambridge} \\
\textit{Cambridge 2} & 2,536 & 2,064,114 & 3.89d & 1s & 5,996 & 42 &\cite{cam-leguay06} \\
\textit{Intel} & 112 & 2,448,720 & 4.15d & 20s & 107 & 1,326 &\cite{cambridge} \\
\textit{Copenhagen Bluetooth} & 671 & 458,920 & 28.0d & 20s & 13,363 & 2&\cite{stopczynski2014measuring} \\
\textit{Kenya} & 52 & 2,070 & 2.54d & 1h & 43 & 26&\cite{kenya} \\
\textit{Diary} & 49 & 2,143 & 4.28y & 1d & 345 & 4&\cite{read} \\
\textit{Prostitution} & 16,730 & 50,632 & 6.00y & 1d & 39,044 & 1&\cite{prostitution} \\
\textit{St Andrews} & 25 & 408,996 & 74d & 1s & 139 & 379 &\cite{dsn} \\
\textit{WiFi} & 18,719 & 9,094,619 & 83.7d & 5m & 884,800 & 6 &\cite{wifi} \\
\textit{Facebook} & 45,813 & 855,542 & 4.28y & 1s & 183,412 & 1&\cite{mislove} \\
\textit{Messages} & 35,624 & 489,653 & 8.27y & 1s & 94,768 & 2&\cite{karimi}  \\
\textit{Forum} & 7,084 & 1,429,573 & 8.61y & 1s & 70,942 & 2&\cite{karimi} \\
\textit{Dating} & 29,341 & 529,890 & 1.15y & 1s & 74,561 & 2&\cite{pok} \\
\textit{Copenhagen Calls} & 483 & 10,545 & 28.0d & 1s& 271 & 6&\cite{stopczynski2014measuring} \\
\textit{Copenhagen SMS} & 533 & 30,380 & 21.6d &1s & 320 & 12 &\cite{stopczynski2014measuring} \\
\textit{E-mail 1} & 57,194 & 444,160 & 112d & 1s & 92,442 & 1&\cite{ebel} \\
\textit{E-mail 2} & 3,188 & 309,125 & 81d & 1s & 16,220 & 3&\cite{eckmann} \\
\textit{E-mail 3} & 986 & 332,334 & 1.52y & 1s & 9,474 & 3&\cite{eml3} \\
\textit{E-mail 4} & 167 & 82,927 & 271d & 1s & 1,830 & 4&\cite{radek} \\
\textit{College} & 1,899 & 59,835 & 193d & 1s & 8,608 & 2&\cite{college} \\
\end{tabular}
\end{ruledtabular}
\end{table*}

\subsection{Empirical networks}

As motivated in the Introduction, we base our study on empirical temporal networks. All networks that we study record contacts between people and falls into two classes---human proximity networks and communication networks. Proximity networks are, of course, most relevant for epidemic studies, but communication networks can serve as a reference (and it is interesting to see how general results are over the two classes). The data sets consist of anonymized lists of two id-numbers in contact and the time since the beginning of the contact.

Many of the proximity data sets we use come from the Sociopatterns project~\cite{sociopatterns}. These data sets were gathered by people wearing radio-frequency identification (RFID) sensors that detect proximity between 1 and 1.5 m. One such datasets comes from a conference, Hypertext 2009, (\textit{Conference 1})~\cite{conference}, another two from a primary school (\textit{Primary School})~\cite{school} and five from a high school (\textit{High School})~\cite{hschool}, a third from a hospital (\textit{Hospital})~\cite{hospital}, a fourth set of five data sets from an art gallery (\textit{Gallery})~\cite{gallery}, a fifth from a workplace (\textit{Office})~\cite{office}, and a sixth from members of five families in rural \textit{Kenya}~\cite{kenya}. 
The \textit{Gallery} data sets consist of several days where we use the first five.

In addition to data gathered by RFID sensors, we also use data from the longer-range (around 10m) Bluetooth channel. The \textit{Cambridge 1}~\cite{cambridge} and \textit{2}~\cite{cam-leguay06} datasets were measured by the Bluetooth channel of sensors (iMotes) worn by people in and around Cambridge UK. \textit{St Andrews}~\cite{dsn}, \textit{Conference 2}~\cite{cambridge} and \textit{Intel}~\cite{cambridge} are similar data sets tracing contacts at, respectively, the University of St Andrews, the conference Infocom 2006 and the Intel research lab in Cambridge UK. The \textit{Reality}~\cite{reality} and \textit{Copenhagen Bluetooth}~\cite{stopczynski2014measuring} data sets also come from Bluetooth data, but from smartphones worn by university students. In the \textit{Romania} data, the WiFi channel of smartphones was used to log the proximity between university students~\cite{roman}, whereas the \textit{WiFi} dataset links students of a Chinese university that are logged onto the same WiFi router. For the \textit{Diary} data set, a group of colleagues and their family members were self-recording their contacts~\cite{read}.
Our final proximity data, the \textit{Prostitution} network, comes from from self-reported sexual contacts between female sex-workers and their male sex-buyers~\cite{prostitution}. This is a special form of proximity network since contacts represent more than just proximity.

Among the data sets from electronic communication, \textit{Facebook} comes from the wall posts at the social media platform Facebook~\cite{mislove}. \textit{College} is based on communication at a Facebook-like service~\cite{college}. \textit{Dating} shows interactions at an early Internet dating website~\cite{pok}. \textit{Messages} and \textit{Forum} are similar records of interaction at a film community~\cite{karimi}. \textit{Copenhagen Calls} and \textit{Copenhagen SMS} consist of phone calls and text messages gathered in the same experiment as \textit{Copenhagen Bluetooth}~\cite{stopczynski2014measuring}. Finally, we use four data sets of e-mail communication. One, \textit{E-mail 1}, recording all e-mails to and from a group of accounts~\cite{ebel}. The other three, \textit{E-mail 2}~\cite{eckmann}, \textit{3}~\cite{eml3}, and \textit{4}~\cite{radek} recording e-mails within a set of accounts.

We list basic statistics---sizes, sampling durations, etc.---of all the data sets in Table~\ref{tab:data}.

\subsection{Static network descriptors}

To gain further insight into the network structures promoting the objective measures, we correlate the objective measures with quantities describing the position of a node in the static networks---\text. Since many of our networks are fragmented into components, we restrict ourselves to measures that are well defined for disconnected networks. Otherwise, in our selection, we strive to cover as many different aspects of node importance as we can.

\subsubsection{Degree}
\textit{Degree} is simply the number of neighbors of a node. It usually presented as the simplest measure of centrality and one of the most discussed structural predictors of importance with respect to disease spreading~\cite{newman:book}. (Centrality is a class of measures of a node's position in a network that try to capture what a ``central'' node is---i.e.\ ultimately centrality is not more well-defined than the vernacular word.) It is also a local measure in the sense that a node is able to estimate its degree, which could be practical when evaluating sentinel surveillance in real networks.

\subsubsection{Subgraph centrality}

\textit{Subgraph centrality} is based on the number of closed walks a node is a member of. (A walk is a path that could be overlapping with itself.) The number of paths from node $i$ to itself is given by ${\bf A}^k_{ii}$, where ${\bf A}$ is the adjacency matrix. Ref.~\cite{subgraphcent} argues that the best way to weigh paths of different lengths together is through the formula
\begin{equation}
C_S(i)=\sum_k \frac{{\bf A}^k_{ii}}{k!} .
\end{equation}

\subsubsection{Component size}

As mentioned, several of the data sets are fragmented (even though the largest connected component dominates components of other sizes). In the limit of high transmission probabilities, all nodes in the component of the infection seed will be infected. In such a case it would make sense to place a sentinel in the largest component (where the disease most likely starts).

\subsubsection{Harmonic closeness}

Closeness centrality builds on the assumption that a node that has, on average, short distances to other nodes is central~\cite{sab:clo}. Here, the distance $d(i,j)$ between nodes $i$ and $j$ is the number of links in the shortest paths between the nodes. The classical measure of closeness centrality of a node $i$ is the reciprocal average distance between $i$ and all other nodes. In a fragmented network, for all nodes, there will be some other node that it does not have a path to, meaning that the closeness centrality is ill defined. (Assigning the distance infinity to disconnected pairs would give the closeness centrality zero for all nodes.) A remedy for this is to, instead of measuring the reciprocal average of distances, measure the average reciprocal distance~\cite{holme_ghoshal}
\begin{equation}
C_C(i,G) = \frac{1}{N}\sum_{j\neq i} \frac{1}{d(i,j)} ,
\end{equation}
where $d^{-1}(i,j) = 0$ if $i$ and $j$ are disconnected. We call this the \textit{harmonic closeness} by analogy to the harmonic mean.

\subsubsection{Harmonic vitality}

Vitality measures are a class of network descriptor that capture the impact of deleting a node on the structure of the entire network~\cite{corley,koschutzki2005centrality}. Specifically, we measure the harmonic closeness vitality, or \textit{harmonic vitality}, for short. This is the change of the sum of reciprocal distances of the graph (thus, by analogy to the harmonic closeness, well-defined even for disconnected graphs).
\begin{equation}
C_V(i,G) = \frac{\sum_{j\in G} C_C(j,G)}{\sum_{j\in G\setminus \{i\}} C_C(j,G\setminus \{i\})} .
\end{equation}
Here the denominator concerns the graph $G$ with the node $i$ deleted. If deleting $i$ breaks many shortest paths, then $C_C(i)$ decreases, and thus $C_V(i)$ increases. A node whose removal disrupts many shortest paths would thus score high in harmonic vitality.

\subsubsection{Coreness}

Our sixth structural descriptor is \textit{coreness}. This measure comes out of a procedure called $k$-core decomposition. First, remove all nodes with degree $k=1$. If this would create new nodes with degree one, delete them too. Repeat this until there are no nodes of degree one. Then, repeat the above steps for larger $k$-values. The coreness of a node is the last level when it is present in the network during this process~\cite{harary}.

\subsection{Temporal network descriptors}

\subsubsection{Degree}

Like for the static networks, in the temporal networks we measure the degree of the nodes. To be precise, we define the degree as the number of distinct other nodes a node in contact with within the data set.

\subsubsection{Strength}

Strength is the total number of contacts a node has participated in throughout the data set. Unlike degree, it takes the number of encounters into account.

\subsubsection{Up- and downstream component sizes}

Temporal networks, in general, tend to be more disconnected than static networks. For node $i$ to be connected to $j$ in a temporal networks there has to be a time-respecting path from $i$ to $j$---i.e.\ a sequence of contacts increasing in time that (if time is projected out) is a path from $i$ to $j$~\cite{holme_modern,masuda_lambiotte}. Thus two interesting quantities---corresponding to the component sizes of static networks---are the fraction of nodes reachable from a node by time-respecting paths forward (downstream component size) and backward in time (upstream component size)~\cite{holme2005}.

\subsubsection{Temporal statistics}

If a node only exist in the very early stage of the data, the sentinel will likely not be active by the time the outbreak happens. If a node is active only at the end of the data set, it would also be too late to discover an outbreak early. For these reasons, we measure statistics of the times of the contacts of a node. We measure the \textit{average time} of all contacts a node participate in; the \textit{first time} of a contact (i.e.\ when the node enters the data set); and the \textit{duration} of the presence of a node in the data (the time between the first and last contact it participates in).

\subsection{Modified Kendall's $\tau$-coefficient}

We use a version of the Kendall $\tau$ coefficient~\cite{kendall} to elucidate both the correlations between the three objective measures, and between the objective measures and network structural descriptors. In its basic form, the Kendall $\tau$ measures the difference between the number of concordant (with a positive slope between them) and discordant pairs relative to all pairs. There are a few different versions that handle of ties in different ways. We count a pair of points whose errorbars overlap as a tie and calculate
\begin{equation}
\tau=\frac{n_c - n_d}{n_c + n_d + n_t},
\end{equation}
where $n_c$ is the number of concordant pairs, $n_d$ is the number of discordant pairs, and $n_t$ is the number of ties.

\begin{figure*}
\includegraphics[width=\textwidth]{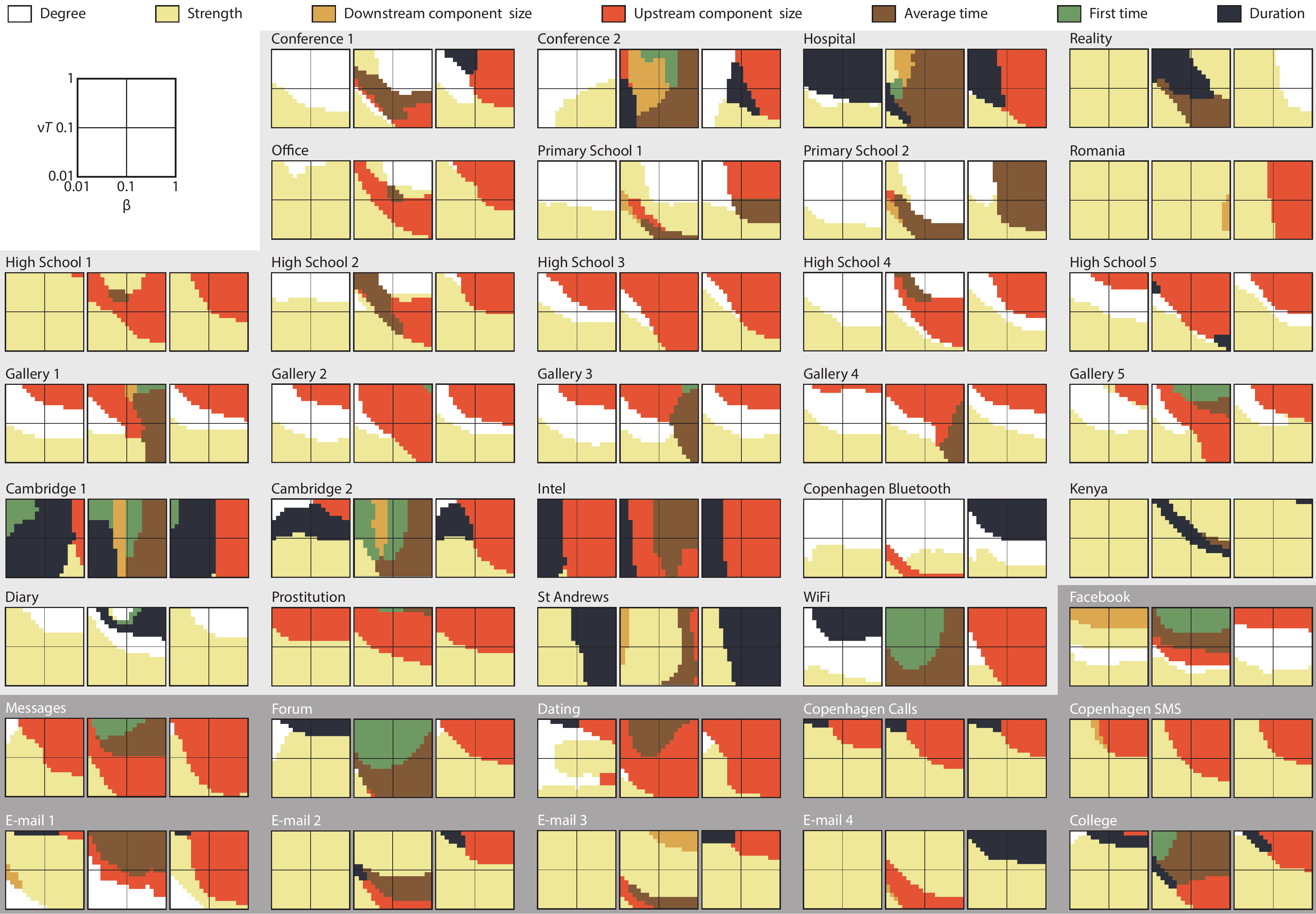}
\caption{(Color online) For every $\beta$ value, this figure shows the strongest correlation between the three objective measures and various measures of network position for the temporal networks. For each data set, the upper panel shows the results for $t_x$, the middle panel $t_d$ and the lower panel $f_d$. The lighter background shows data sets of human proximity while the darker background are based on human communication data. The axes are logarithmic.}
\label{fig:temp_struct}
\end{figure*}

\section{Results}

\subsection{Correlation between the objective measures}

We start investigating the correlation between the three objective measures throughout the parameter space of the SIR model for all our data sets.

\subsubsection{Static networks}

We use the time to detection and extinction as our baseline and compare the other two objective measures with that. In Fig.~\ref{fig:mcorr_stat}, we plot the $\tau$ coefficient between $t_x$ and $t_d$, and $t_x$ and $f_d$ respectively. We find that for low enough values of $\beta$, the $\tau$ for all objective measures coincide. For very low $\beta$ the disease just dies out immediately, so the measures are trivially equal---all nodes would be as good sentinels in all three aspects. For slightly larger $\beta$---for most data sets $0.01 < \beta <0.1$---both $\tau(t_x,t_d)$ and $\tau(t_x,f_d)$ are negative. This is a region where outbreaks typically die out early. For a node to have low $t_x$, it needs to be where outbreaks are likely survive, at least for a while. This translates to a large $f_d$, while for $t_d$, it would be beneficial to be as central as possible.

If there are no extinction events at all, $t_x$ and $t_d$ are the same. For this reason, it is no surprise that, for most of the data sets, $\tau(t_x,t_d)$ becomes strongly positively correlated for large $\beta$-values. The $\tau(t_x,f_d)$ correlation is negative (of a similar magnitude), meaning that for most data sets the different methods would rank the possible sentinels in the same order. For some of the data sets, however, the correlation does never become positive even for large $\beta$ values (like \textit{Copenhagen Calls} and \textit{Copenhagen SMS}). These networks are the most fragmented ones meaning that one sentinel unlikely would detect the outbreak (since it probably happens in another component). This makes $t_x$ rank the important nodes in a similar way to $f_d$, but since diseases that do reach a sentinel does it faster in a small component than a large, $t_x$ and $t_d$ becomes anti correlated.

\subsubsection{Temporal networks}

In Fig.~\ref{fig:mcorr_temp}, we perform the same analysis as in the previous section but for static networks. The picture is to some extent similar, but also much richer. Just as for the case of static networks, $\tau(t_x,f_d)$ is always non-positive, meaning the time to detection or extinction ranks the nodes in a way positively correlated with the frequency of detection. Furthermore, like the static networks $\tau(t_x,t_d)$ can be both positively and negatively correlated. This means that there are regions where $t_d$ ranks the nodes in the opposite way than the $t_x$. These regions of negative $\tau(t_x,t_d)$ occur for low $\beta$ and $\nu$. For some data sets---for example the \textit{Gallery} data sets, \textit{Dating}, \textit{Copenhagen calls} and \textit{Copenhagen SMS}---the correlations are negative throughout the parameter space.

Among the data sets with a qualitative difference between the static and temporal representations, we find \textit{Prostitution} and \textit{E-mail 1} that both have strongly positive values of $\tau(t_x,t_d)$ for large $\beta$-values in the static networks but moderately negative values for temporal networks.

\subsection{Correlation between objective measures and structural descriptors}

In this section, we take a look at how network structure affect our objective measures.

\subsubsection{Static networks}

In Fig.~\ref{fig:office_stat_corr}, we show the correlation between our three objective measures and the structural descriptors as a function of $\beta$ for the \textit{Office} data set. Panel (a) shows the results for the time to detection or extinction. There is a negative correlation between this measure and traditional centrality measures like degree or subgraph centrality. This is because $t_x$ is a quantity one wants to minimize to find the optimal sentinel, whereas for all the structural descriptors, a large value means that a node is a candidate sentinel node. We see that degree and subgraph centrality are the two quantities that best predict the optimal sentinel location, while coreness is also close (at around $-0.65$). This in line with research showing that certain biological problems are better determined by degree than more elaborate centrality measures~\cite{klemm}. Over all, the $\tau$ curves are rather flat. This is partly explained by $\tau$ being a rank correlation coefficient---if the rankings do not change (even if the objective measures do), then neither do the $\tau$ values.

For $t_d$ (Fig.~\ref{fig:office_stat_corr}(b)), most curves change behavior around $\beta=0.2$. This is the region when larger outbreaks could happen, so one can understand there is a transition to a situation similar to $t_x$ (Fig.~\ref{fig:office_stat_corr}(a)). $f_d$ (Fig.~\ref{fig:office_stat_corr}(c)) shows a similar behavior to $t_d$ in that the curves start changing order, and what was a correlation at low $\beta$ becomes an anti-correlation at high $\beta$. This anti-correlation is a special feature of this particular data set, perhaps due to its pronounced community structure. Nodes of degree 0, 1 and 2 have a strictly increasing values of $f_d$, but for some of the high degree nodes (that all have $f_d$ close to one) the ordering gets anti-correlated with degree which makes Kendall's $\tau$ negative. Since rank-based correlations are more principled for skewedly distributed quantities common in networks, we keep them. We currently investigate what creates these unintuitive anti-correlations among the high degree nodes in this data set.

Next, we proceed with an analysis of all data sets. We summarize plots like Fig.~\ref{fig:office_stat_corr} by the structural descriptor with the largest magnitude of the correlation $|\tau |$. See Fig.~\ref{fig:mcorr_stat}. We can see, that there is not one structural quantity that uniquely determines the ranking of nodes, there is not even one that dominates over the range of $\beta$ that we investigate. Furthermore, there are some striking patterns:
\begin{enumerate}[label=(\roman*)]
\item Degree is the strongest structural determinant of all objective measures at low $\beta$-values. This is consistent to Ref.~\cite{holme3f}.
\item Component size only occurs for large $\beta$. In the limit of large $\beta$, $f_d$ is only determined by component size (if we would extend the analysis to even larger $\beta$ subgraph centrality would have the strongest correlation for the frequency of detection).
\item Harmonic vitality is relatively better as a structural descriptor for $t_d$, less so for $t_x$ and $f_d$. $t_x$ and $f_d$ capture the ability of detecting an outbreak before it dies, so for these quantities one can imagine more fundamental quantities like degree and the component size are more important.
\item Subgraph centrality often shows the strongest correlation for intermediate values of $\beta$. This is interesting, but difficult to explain since the rationale of subgraph centrality builds on cycle counts and there is no direct process involving cycles in the SIR model.
\item Harmonic closeness rarely gives the strongest correlation. If it does, it is usually succeeded by coreness and the data set is typically rather large.
\item Datasets from the same category can give different results. Perhaps \textit{College} and \textit{Facebook} are the most conspicuous example. In general, however, similar data sets give similar results.
\end{enumerate}
The final observation could be extended. We see that, as $\beta$ increases, one color tends to follow another. This is summarized in Fig.~\ref{fig:transitions} where we show transition graphs of the different structural descriptors such that the size corresponds to their frequency in Fig.~\ref{fig:mcorr_stat}, and the size of the arrows show how often one structural descriptor is succeeded by another as $\beta$ is increased. For $t_x$, the degree and subgraph centrality are the most important structural descriptors and the former is usually succeeded by the former. For $t_d$, there is a common peculiar sequence of: degree, subgraph centrality, coreness component size and harmonic vitality that is manifested as the peripheral, clock-wise path of Fig.~\ref{fig:transitions}(b). Finally, $f_d$ is similar to $t_x$ except that there is a rather common transition from degree to coreness, and harmonic vitality is, relatively speaking, a more important descriptor.

\subsubsection{Temporal networks}

In Fig.~\ref{fig:mcorr_temp}, we show the figure for temporal networks corresponding to Fig.~\ref{fig:mcorr_stat}. Just like the static case, even though every data set and objective measure is unique, we can make some interesting observations.
\begin{enumerate}[label=(\roman*)]
\item Strength is most important for small $\nu$ and $\beta$. This is analogous to degree dominating the static network at small parameter values.
\item Upstream component size dominates at large $\nu$ and $\beta$. This is analogous to the component size of static networks. Since temporal networks tend to be more fragmented than static ones~\cite{holme2005}, this dominance at large outbreak sizes should be even more pronounced for temporal networks.
\item Most of the variation happens in the direction of larger $\nu$ and $\beta$. In this direction, strength is succeeded by degree which is succeeded by upstream component size.
\item Like the static case, and the analysis of Figs.~\ref{fig:stat_struct} and \ref{fig:temp_struct}, $t_x$ and $f_d$ are qualitatively similar compared to $t_d$.
\item Temporal quantities, such as the average and first times of a node's contacts are commonly the strongest predictors of $t_d$.
\item When a temporal quantity is the strongest predictor of $t_x$ and $f_d$ it is usually the duration. It is understandable that this has little influence on $t_d$, since the ability to be infected at all matters for these measures, a long duration is beneficial since it covers many starting times of the outbreak.
\item Similar to the static case, most categories of data sets give consistent results, but some differ much (\textit{Facebook} and \textit{College} is yet again a good example).
\end{enumerate}
The bigger picture these observations paint is that, for our problem, the temporal and static networks behaves rather similarly, meaning that the structures in time do not matter so much for our objective measures. At the same time, there is not only one dominant measure for all the data sets. Rather are there several structural descriptors that correlates most strongly with the objective measures depending on $\nu$ and $\beta$.

\section{Summary and conclusions}

In this paper, we have investigated three different objective measures for optimizing sentinel surveillance: the time to detection or extinction, the time to detection (given that the detection happens) and the frequency of detection. Each of these measures corresponds to a public health scenario---the time to detection or extinction is most interesting to minimize of one want to halt the outbreak as quickly as possible, the frequency of detection is most interesting if one wants to monitor the epidemic status as accurately as possible. The time to detection is interesting if a one want to detect the outbreak early (or else it is not important), which could be the case if manufacturing new vaccine is relatively time consuming. We investigate these cases for 38 temporal network data sets and static networks derived from the temporal networks.

Our most important finding is that, for some regions of parameter space, our three objective measures can rank nodes very differently. This comes from the fact that SIR outbreaks have a large chance of dying out in the very early phase~\cite{janson:law}, but once they get going they follow a deterministic path. For this reason, it is thus important to be aware of what scenario one is investigating when addressing the sentinel surveillance problem.

Another conclusion is that for this problem, static and temporal networks behave reasonably similar (meaning that the temporal effects do not matter so much). Naturally, some of the temporal networks respond differently than the static ones, but compared to, e.g., the outbreak sizes or time to extinction~\cite{holme_tempdis,feff,bansal}, differences are small.

Among the structural descriptors of network position, there is no particular one that dominates throughout the parameter space. Rather, local quantities like degree or strength (for the temporal networks) have a higher predictive power at low parameter values (small outbreaks). For larger parameter values, descriptors capturing the number of nodes reachable from a specific node correlates most with the objective measures rankings. Also in this sense, the static network quantities dominates the temporal ones, which is in contrast to previous observations (e.g.\ Refs.~\cite{holme_tempdis,feff,bansal}).

For the future, we anticipate work on the problem of optimizing sentinel surveillance. An obvious continuation of this work would be to establish the differences between the objective metrics in static network models. To do the same in temporal networks would also be interesting, although more challenging given the large number of imaginable structures. Yet an open problem is how to distribute sentinels if they are more than one. It is known that they should be relatively far away~\cite{holme3f}, but more precisely where?

\begin{acknowledgments}
We thank Sune Lehmann for providing the Copenhagen data sets. This work was supported by JSPS KAKENHI Grant Number JP 18H01655.
\end{acknowledgments}

\bibliographystyle{abbrv}
\bibliography{exact}

\end{document}